\documentclass[11pt,aip,reprint,jcp,longbibliography]{revtex4-1}
\usepackage[utf8]{inputenc}
\usepackage[a4paper]{geometry}
\usepackage{amsmath}
\usepackage{bm}
\usepackage{amssymb}
\usepackage{graphicx}
\usepackage{esint}

\makeatletter

\providecommand{\tabularnewline}{\\}

\usepackage[inline]{enumitem}
\usepackage{xcolor}
\newcommand{\red}{\textcolor{black}}

\makeatother

\begin{document}

\title{Assessing the correctness of pressure correction to solvation theories
in the study of electron transfer reactions}

\author{Tzu-Yao Hsu}

\affiliation{Sorbonne Université, CNRS, Physico-Chimie des Électrolytes et Nanosystèmes
Interfaciaux, PHENIX, F-75005 Paris, France}

\author{Guillaume Jeanmairet}
\email[]{guillaume.jeanmairet@sorbonne-universite.fr}
\affiliation{Sorbonne Université, CNRS, Physico-Chimie des Électrolytes et Nanosystèmes
Interfaciaux, PHENIX, F-75005 Paris, France}
\affiliation{Réseau sur le Stockage Électrochimique de l'Énergie (RS2E), FR CNRS
3459, 80039 Amiens Cedex, France}

\begin{abstract}
Liquid states theories have emerged as a numerically efficient alternative
to costly molecular dynamics simulations of electron transfer reactions
in solution. In a recent paper {[}\textit{Chem. Sci.}, 2019, 10, 2130{]},
we introduced the framework to compute energy gap, free energy profile
and reorganization free energy using molecular density functional
theory. However, this technique, as other molecular liquid state theories,
overestimates the bulk pressure of the fluid. Because of the too high pressure,
the predicted free energy  is dramatically exaggerated. Several attempts were made to fix this issue,
either based on simple a posteriori correction or by introducing bridge terms. By studying two
model half reactions in water, $\text{Cl}\rightarrow\text{Cl}^{+}$
and $\text{Cl}\rightarrow\text{Cl}^{-}$, we assess the correctness
of these two types of corrections to study electron transfer reactions.
We found that a posteriori correction, because it violates the functional
principle, leads to an inconsistency in the definition of the reorganization
free energy and should not be used to study electron transfer reactions. The
bridge approach, because it is theoretically well grounded, is perfectly
suitable for this type of systems.
\end{abstract}
\maketitle

\section{Introduction\label{sec:Introduction}}

The widely accepted theory to describe an electron transfer reaction
 in solution was proposed by Marcus in 1956 \cite{marcus_theory_1956}.
Assuming the solvent can be described by a polarization field that
responds linearly to the electric field generated by the solute, he
derived an expression linking the activation free energy to the reaction
free energy. This expression depends on an unique parameter: the reorganization
energy. \red{Reorganization energy}  measures the cost to distort the solvent from the
configuration in equilibrium with the reactant to the one in equilibrium
with the product, without transferring electron. The polarization varies continuously between the equilibrium
values of reactant and product. By computing the free energy of both
states for out-of equilibrium polarizations, one can plot the free
energies as a function of an appropriate reaction coordinate. This
gives rise to the famous Marcus two parabola picture for the free
energy profiles.

From a simulation point of view, \red{electron transfer reactions} have been mostly investigated
using molecular dynamics (MD). Warshel proposed to use the energy
gap  as the relevant microscopic reaction coordinate \cite{warshel_dynamics_1982}.
When the probability distribution of the \red{energy gap} is Gaussian, the \red{fee energy profiles} are
parabolic with identical curvatures, as predicted by Marcus\cite{ferrario_redox_2006}.
The Gaussian behavior has been verified in several studies but some
systems deviate from the Marcus picture\cite{hartnig_molecular_2001,blumberger_cuaq+/cuaq2+_2008,vuilleumier_extension_2012}.
Resort to numerical simulations of \red{electron transfer reactions} is essential to account for
such cases, unfortunately computing the \red{free energy profiles} with MD remains costly
and require the use of biased sampling techniques\cite{king_investigation_1990,vandevondele_density_2006}. 

We recently proposed an alternative way to compute \red{free energy profiles} and reorganization
free energies of \red{electron transfer reactions} based on molecular Density Functional Theory
(mDFT)\cite{borgis_molecular_2012,jeanmairet_molecular_2019}. Because
it allows to compute directly the solvation free energy through functional
minimization, it is computationally more efficient than MD. Moreover,
mDFT allows a fine description of solvation at the molecular level
and a good agreement with MD was found for the $\text{Cl}\rightarrow\text{Cl}^{+}$
and the $\text{Cl}\rightarrow\text{Cl}^{-}$ \red{electron transfer reactions}. The first one is
found to follow Marcus theory while the second one is not.

In previous work, the hyper-netted chain (HNC) approximation was used to describe the
solute-solvent correlations\cite{ding_efficient_2017}, neglecting
the so-called bridge functional. However, because this functional is a
Taylor expansion truncated at second-order, it drastically overestimates
the free energy per solvent molecule at gas density\cite{rickayzen_integral_1984,borgis_simple_2020,sergiievskyi_fast_2014}.
This increases considerably the cost to create a solvent cavity and
thus the predicted solvation free energy. The reference interaction
site model and its 3D variant (3D-RISM), another liquid state theory
that have also been extensively used to tackle \red{electron transfer reactions}\cite{yamaguchi_nonequilibrium_2020,chong_free_1996,chong_molecular_1995},
suffer from the same problem\cite{sergiievskyi_fast_2014,gillespie_restoring_2014,sergiievskyi_solvation_2015,misin_predicting_2016,misin_salting-out_2016}.
Because it is a long-standing problem, several attempts were made to fix
it. By noticing that the gas phase free energy is directly linked
to the pressure, simple a-posteriori corrections were proposed \cite{luukkonen_hydration_2020,sergiievskyi_fast_2014,robert_pressure_2020}
\cite{palmer_towards_2010,ratkova_accurate_2010,misin_communication_2015,sergiievskyi_solvation_2015}.
Essentially, they consist in evaluating the partial molar volume (PMV)
created by the solute and applying a free energy correction proportional
to it. 

A more rigorous approach is to fix the defect of the functional by
introducing an approximate bridge term, to go beyond second order.
Several models of bridge functional have been proposed, either based
on hard sphere theory\cite{oettel_integral_2005,levesque_scalar_2012,fu_fast_2014}
or weighted density approximation (WDA) \cite{jeanmairet_molecular_2013,jeanmairet_molecular_2015,borgis_simple_2020}. 

In this paper, we evaluate the impact of those two types of correction
on the \red{free energy profiles} and reorganization free energies predicted for two model
\red{electron transfer reactions}: $\text{Cl}\rightarrow\text{Cl}^{+}$ and $\text{Cl}^{-}\rightarrow\text{Cl}$.
For a posteriori correction we selected the widely used \red{pressure correction (PC)}  \cite{sergiievskyi_fast_2014,sergiievskyi_solvation_2015}
while the bridge functional is the recently proposed angular independent
WDA functional \cite{borgis_simple_2020}.

\section{Theory\label{sec:Theory}}

We remind here some basics of mDFT which are essential for the comprehension
of this paper. Thorough descriptions can be found in previous reports\cite{luukkonen_hydration_2020,borgis_molecular_2012,jeanmairet_molecular_2019}.

\label{secMDFT} In mDFT, solvent and solute molecules are described
by rigid models interacting through classical forcefield. The solvation
free energy of the solute can be computed by minimizing the solvent
functional with respect to the spatially and orientationally dependent
solvent density $\rho(\bm{r},\bm{\Omega})$. We introduce the following
notation $\mbox{\ensuremath{\bm{x}\equiv\left(\bm{r},\bm{\Omega}\right)}}$
for concision. The functional to be minimized is 
\begin{equation}
F[\rho(\bm{x})]=F_{\text{id}}[\rho(\bm{x})]+F_{\text{ext}}[\rho(\bm{x})]+F_{\text{exc}}[\rho(\bm{x})].\label{F=00003DFid+Fexc+Fext}
\end{equation}

Its minimum is reached for the equilibrium solvent density, $\rho_{\text{eq}}$.
In equation \ref{F=00003DFid+Fexc+Fext}, $F_{\text{id}}$ corresponds
to the entropy of the non-interacting liquid

\begin{equation}
F_{\text{id}}[\rho]=k_{B}T\int\left[\rho\left(\bm{x}\right)\ln\left(\frac{\rho\left(\bm{x}\right)}{\rho_{\text{b}}}\right)-\rho\left(\bm{x}\right)+\rho_{\text{b}}\right]d\bm{x}\label{eq:Fid}
\end{equation}
where $k_{B}$ is the Boltzmann constant, $\rho_{{\it \text{b}}}$
is the homogeneous bulk solvent density and $T$ is the temperature.

The second term of equation \ref{F=00003DFid+Fexc+Fext} is due to
solute-solvent interaction and can be expressed as 
\begin{equation}
F_{\text{ext}}[\rho]=\int\rho\left(\bm{x}\right)\red{V}(\bm{x})d\bm{x}\label{eq:Fext}
\end{equation}
where $\red{V}$ is the the external \red{potential} energy exerted
by the solute.

Solvent-solvent interactions are collected in the last term of equation
\ref{F=00003DFid+Fexc+Fext}, that is the excess functional $F_{\text{exc}}$.
Expanding it around the density $\rho_{\text{{\it \text{b}}}}$ of
the homogeneous fluid taken as a reference, and truncating the expansion
at second order gives rise to the HNC functional. All the higher order
correlations are then swept into the unknown bridge functional $F_{\text{B}}$, 

\begin{eqnarray}
F_{\text{exc}}[\rho] & = & F_{\text{HNC}}[\rho]+F_{{\it \text{B}}}[\rho]\label{eq:Fexc}\\
 & = & -\frac{1}{2}k_{B}T\iiiint\left[\Delta\rho(\bm{r}_{1},\bm{\Omega}_{1})c(\bm{r}_{1}-\bm{r}_{2},\bm{\Omega}_{1},\bm{\Omega}_{2})\right.\nonumber \\
 &  & \left.\times\Delta\rho(\bm{r}_{2},\bm{\Omega}_{2})\right]d\bm{r}_{1}d\bm{\Omega}_{1}d\bm{r}_{2}d\bm{\Omega}_{2}+F_{{\it \text{B}}}[\rho]\nonumber 
\end{eqnarray}
where $c$ is the direct correlation function of homogeneous fluid
at density $\rho_{{\it \text{b}}}$ and $\Delta\rho=\rho-\rho_{\text{b}}$\cite{ding_efficient_2017}.

We evaluate two types of pressure correction in this paper. The PMV
correction takes the following expression\cite{sergiievskyi_fast_2014,sergiievskyi_solvation_2015}
\begin{equation}
F_{\text{B}}^{\text{PMV}}=\rho_{\text{b}}k_{B}T(1-\frac{\rho_{\text{b}}}{2}\hat{c}(k=0))\int\frac{\rho_{\text{eq}}\left(\bm{x}\right)-\rho_{\text{b}}}{\rho_{\text{b}}}d\bm{x}\label{eq:FPMV}
\end{equation}
where $\hat{c}$ denotes the Fourrier transform of the direct correlation function. Since equation \ref{eq:FPMV} does not depend on the density $\rho$,
such a correction does not modify the optimization process.
The equilibrium density remains the same as the one obtained using
the HNC functional.

For the bridge correction, we take the recently proposed spherical
WDA functional \cite{borgis_simple_2020}
\begin{eqnarray}
F_{\text{B}}^{\text{WDA}}[\rho] & = & \frac{k_{B}T}{n_{b}^{3}}\iint a \left[ \int\Delta\rho(\bm{r}^{\prime},\bm{\Omega})\right.\nonumber\label{eq:FWDA}\\
 &  & \left.\times w(\left|\bm{r}-\bm{r}^{\prime}\right|)d\bm{r}^{\prime} \right]^{3}d\bm{\Omega}d\bm{r}
\end{eqnarray}
the value of $a$ is uniquely defined by imposing the correct pressure
\cite{borgis_simple_2020} and $w$ is a Gaussian weighting function
\begin{align}
w(r) & =(2\pi\sigma_{w}^{2})^{-3/2}\exp(-r^{2}/2\sigma_{w}^{2})\label{Gaussian}
\end{align}
with $\sigma_{w}=1\ \textrm{Å}$. 

We now turn our attention to the study of electron transfer half-reaction
of the type $\text{Red}\rightarrow\text{Ox}+e^{-}$, and remind how
it is possible to compute \red{free energy profiles} and reorganization free energies using
mDFT.

We start by introducing a set of external potentials defined as a linear
combination between the external potential of the $\text{Ox}$ (denoted
by 0) and $\text{Red}$ (denoted by 1) states
\begin{equation}
\red{V_{\eta}=V_{0}+\eta(V_{1}-V_{0})}.\label{linearcombine-1}
\end{equation}
where $\eta$ is a real number.

Minimizing the functional in equation \ref{F=00003DFid+Fexc+Fext}
using the external potentials in equation \ref{linearcombine-1} generate
a set of solvent densities $\rho_{\eta}$. While $\rho_{0}$ and $\rho_{1}$
correspond to the equilibrium solvent densities for the $\text{Ox}$
and the $\text{Red}$ states respectively, any other value of $\eta$
defines a solvent density $\rho_{\eta}$ that is out of equilibrium
for both states. It becomes possible to compute the free energy of
an electronic state $\alpha=0\text{ or }1$, solvated in a solvent
density $\rho_{\eta}$, by evaluating the associated functional: $F_{\alpha}[\rho_{\eta}]$.
The definition of the two solvent reorganization free energies comes
out naturally
\begin{equation}
\lambda_{0}=F_{0}[\rho_{1}]-F_{0}[\rho_{0}]\text{ and }\lambda_{1}=F_{1}[\rho_{0}]-F_{1}[\rho_{1}].\label{eq:lambda=00003DDeltaF}
\end{equation}

We then defined the energy gap as
\begin{equation}
\left\langle \Delta E\right\rangle _{\eta}=\int\rho_{\eta}(\bm{x})\left[V_{1}(\bm{x})-V_{0}(\bm{x})\right]d\bm{x}\label{eq:DeltaE_eta}
\end{equation}
and showed that it is an appropriate reaction coordinate since there
is a one to one mapping between the energy gap and the solvent density\cite{jeanmairet_molecular_2019}
\begin{equation}
\eta\leftrightarrow V_{\eta}\leftrightarrow\rho_{\eta}\leftrightarrow\langle\Delta E\rangle_{\eta}.\label{onetoone}
\end{equation}
The free energy of any state is therefore a function of $\langle\Delta E\rangle_{\eta}$
and it is possible to compute the \red{free energy profile} associated to state $\alpha$
as a function of this reaction coordinate
\begin{equation}
F_{\alpha}(\langle\Delta E\rangle_{\eta})\equiv F_{\alpha}[\rho_{\eta}].\label{eq:equationFEC}
\end{equation}

\section{Computational details}

We study two model \red{electron transfer reactions}: $\mbox{\text{Cl}\ensuremath{\rightarrow\text{Cl}^{+}}}$
and $\mbox{\ensuremath{\text{Cl}^{-}\rightarrow\text{Cl}}}$ in water.
The solute is modeled by one Lennard-Jones site, with $\mbox{\ensuremath{\sigma}=4.404\ \textrm{Å}}$
, and $\mbox{\ensuremath{\epsilon}=0.4190\ \text{kJ/mol}}$. We consider
3 oxidation states corresponding to $\text{Cl}^{-}$, $\text{Cl}^{(0)}$
and $\text{Cl}^{+}$. We use a Lennard-Jones cut-off of 10 {Å} with
long range corrections\cite{frenkel_understanding_2002}. Water is
described using the SPC/E model. We generate a series of biased external
potential by varying the atomic charge of the solute between $-1.2\leq q\leq2$
with a step of $1/30$ elementary charge. We use a $\mbox{40\ensuremath{\times}40\ensuremath{\times}40\ \ensuremath{\textrm{Å}^{3}}}$
box with 120$^{3}$ spatial grid points and 196 possible orientations
per spatial point with periodic boundary condition. We use the
type C correction of Hünenberger \cite{kastenholz_computation_2006,kastenholz_computation_2006-1}
due to interaction between the solute and its periodic replica. The
temperature is fixed at 298.15 K. 

\section{Results and discussions}

\subsection{partial molar volume correction}

\label{PMVcorrection} 
\begin{figure}[htbp]
\centering{}%
\begin{tabular}{c}
\includegraphics[width=0.95\columnwidth]{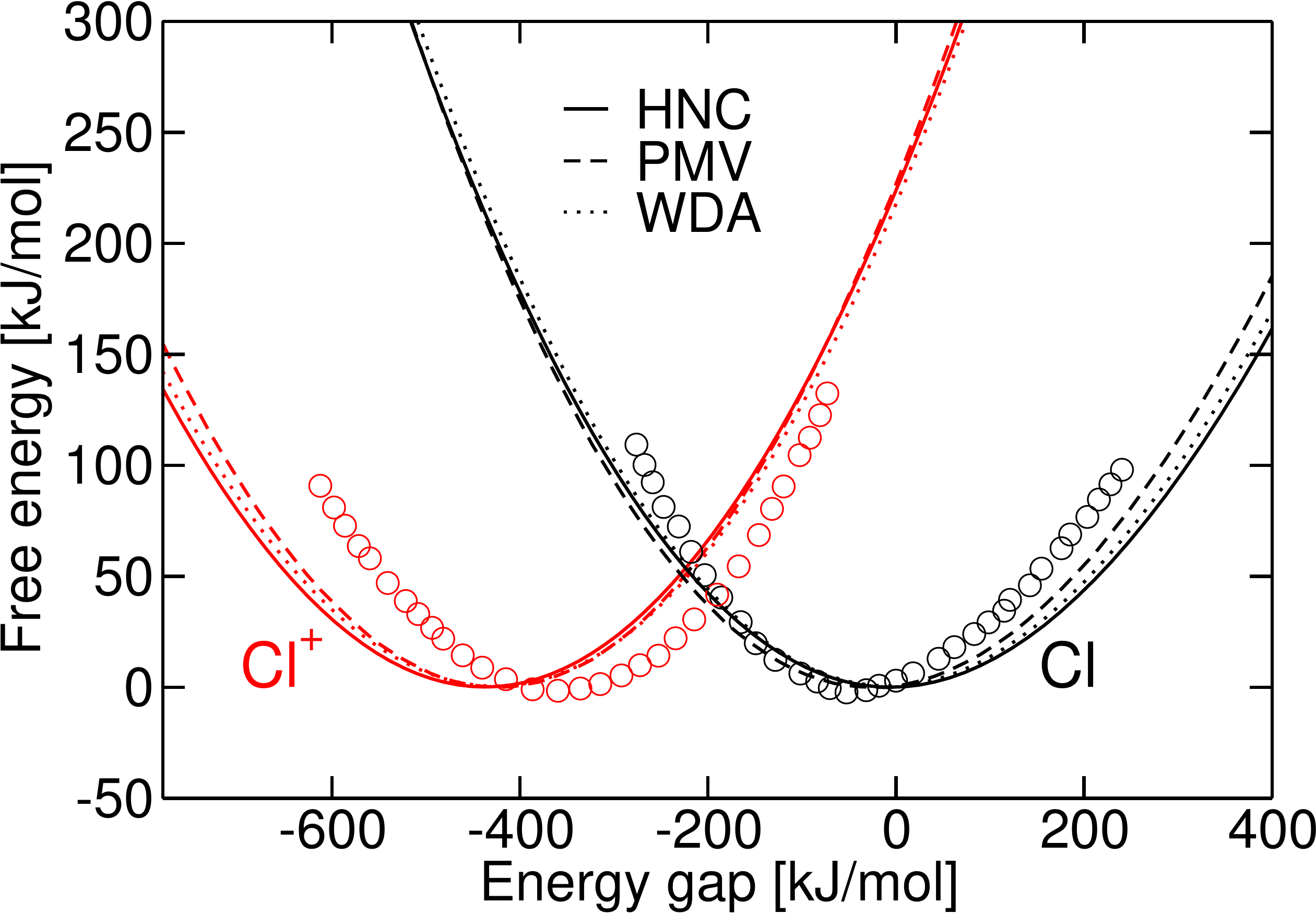}  \tabularnewline\\
\includegraphics[width=0.95\columnwidth]{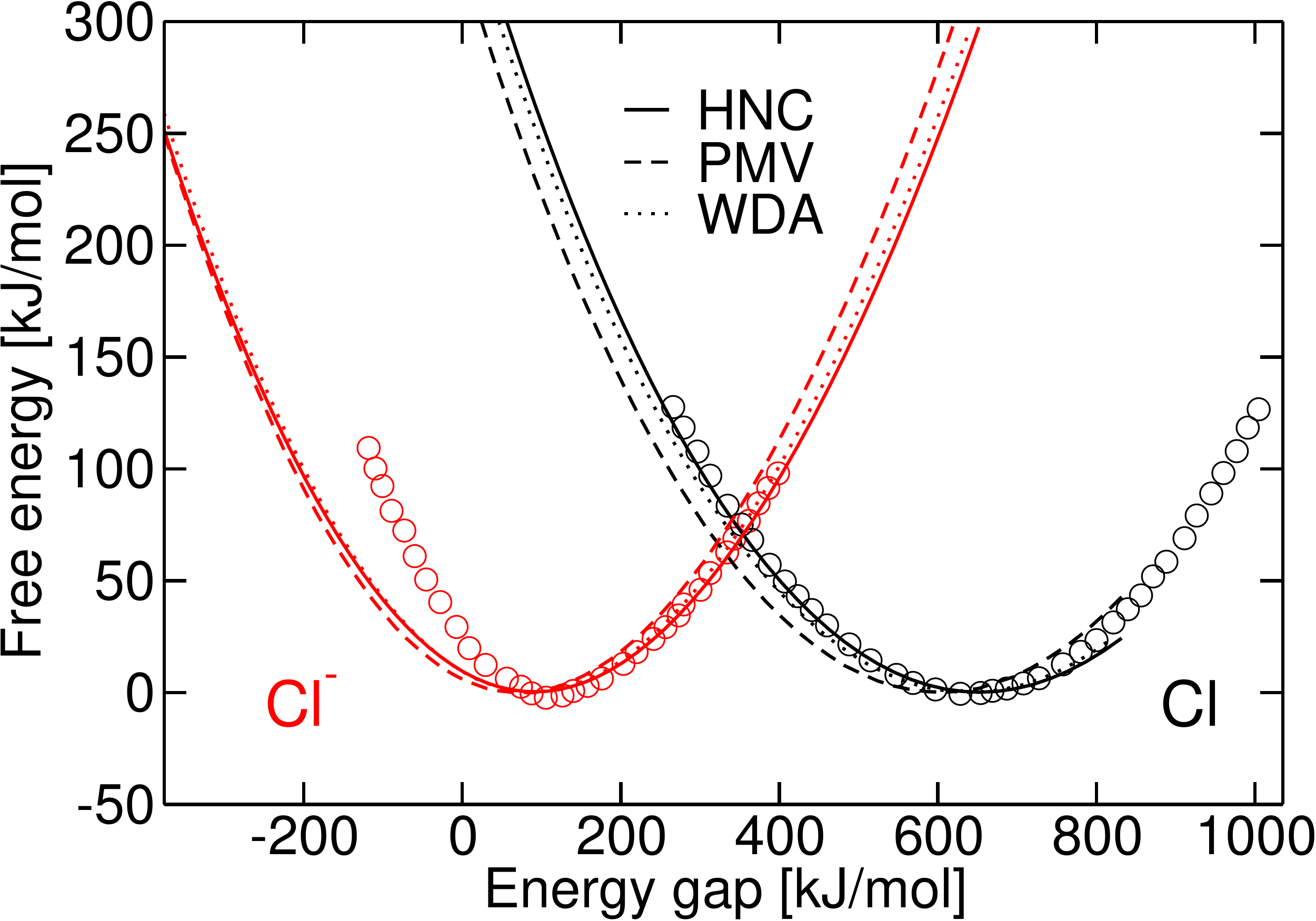}
\end{tabular}\caption{Free energy profiles of $\text{Cl}\rightarrow\text{Cl}^{+}$ and the
$\text{Cl}^{-}\rightarrow\text{Cl}$ \red{electron transfer reactions} computed using the HNC functional
without (full) and with (dashed) the PMV correction and using the
WDA (dotted) bridge functional. The MD data obtained by Hartnig et
al \cite{hartnig_molecular_2001} are the circles.}
\label{fig1: FEC} 
\end{figure}

The free energy profiles of the two half-reactions
calculated using the HNC functional with and without the PMV correction
are displayed in figure \ref{fig1: FEC}. All minima are shifted to
zero to ease the comparison with the curves computed by Hartnig et
al. using MD \cite{hartnig_molecular_2001}. 
\red{The agreement between the free energy profiles obtained using MD and MDFT is satisfactory for the neutral solute. However, important deviations are observed in the case of the cation and the anion. 
This is a know defect of the HNC functional which is not able to properly render the hydrogen bond network of water that is essential for the solvation of ions. This is due to the truncation of the functional expansion at second order. The two types of corrections studied in this paper cannot resolve this issue because of their lack of angular dependency.}

At first glance, including
the PMV correction do not seem to deeply modify the shape of the curve.
However, a closer look reveals that the positions of the minima are
shifted. This is especially true for the ions where the minima are
shifted toward zero. This might appear surprising since the energy gap
defined in equation \ref{eq:DeltaE_eta} solely depends on the density
$\rho_{\eta}$ which is not modified by the PMV correction.

\begin{figure}[htbp]
\centering{}\includegraphics[width=0.95\columnwidth]{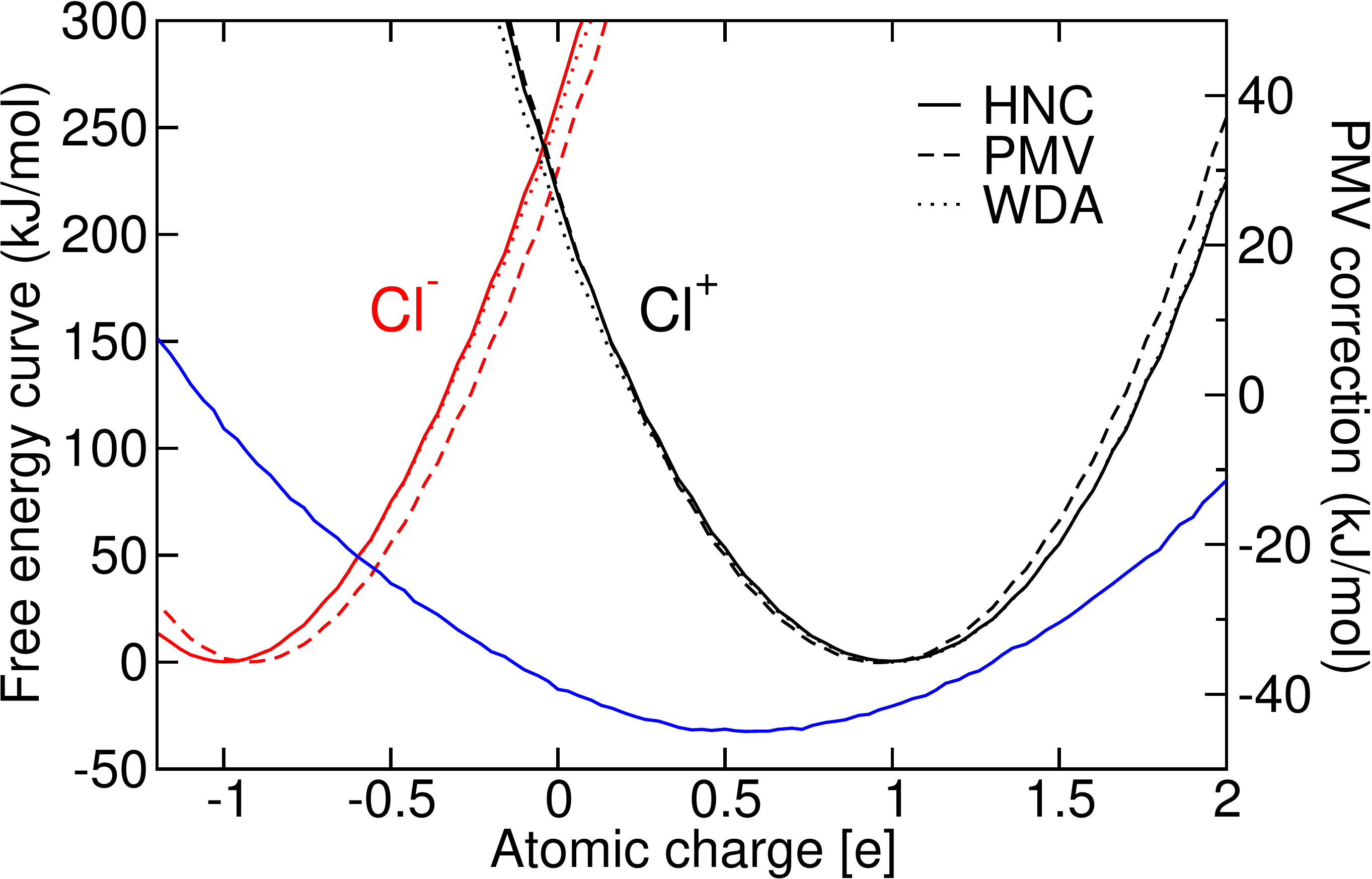}\caption{Free energy of $\text{Cl}^{+}$ and $\text{Cl}^{-}$ computed using
the HNC functional without (full) and with (dashed) the PMV correction
and using the WDA (dotted) bridge functional as a function of the
charge of the fictitious solute. The value of the PMV correction is
also displayed as a function of the charge of the fictitious solute.}
\label{fig2:FEC_and_PMV_as_q} 
\end{figure}

In fact, the shift of the minima is a consequence of a pathological
defect of this correction. This is visible in figure \ref{fig2:FEC_and_PMV_as_q}
where the \red{free energy profiles}  of the ions are depicted as a function of the charge
$q$ of the fictitious solute: $q=\eta$ for the $\text{Cl}\rightarrow\text{Cl}^{+}$
reaction and $q=-\eta$ for the $\text{Cl}^{-}\rightarrow\text{Cl}$
reaction. The minima of the PMV corrected \red{free energy profiles}  no longer correspond
to $q=1$ and $q=-1$. Equivalently, this means that the minimum of
the free energy curve is not reached for the equilibrium solvent density
of the solute. This is a violation of the DFT variational principle
which is not surprising since the PMV correction is not properly integrated
in the classical DFT formalism. It is simply an a posteriori correction
that does not influence the optimization. This \red{might be} acceptable when
the objective is to reproduce some reference solvation free energies
\cite{robert_pressure_2020,sergiievskyi_fast_2014,sergiievskyi_solvation_2015,luukkonen_hydration_2020,misin_communication_2015,misin_predicting_2016}
but it is problematic to study \red{electron transfer reactions}. Indeed, the two ways to compute
the reorganization free energies, either from equation \ref{eq:lambda=00003DDeltaF}
or from the graphical definition, i.e. as the difference between the
value of the free energy of one state at the abscissa corresponding
to the minimum of the second state and the value at its minimum do
not coincide anymore.

Reorganization free energies are given in table \ref{tab}. Note that
there are two sets of values for the neutral chlorine since the value
of $\lambda$ depends on the choice of the second oxidation state.
Solvent reorganization free energies of $\text{Cl}$ and $\text{Cl}^{+}$
computed with the HNC functional are similar indicating that this
\red{electron transfer reaction} is well described using Marcus Theory while  the $\text{Cl}^{-}\rightarrow\text{Cl}$
reaction deviates from Marcus theory. These results are consistent
with MD simulations\cite{hartnig_molecular_2001}. When the PMV correction
is added, both \red{electron transfer reactions} seem to deviate from Marcus theory since the reorganization
free energies of ions differ from the one of the associated neutral
state. Moreover, the reorganization free energies computed using equation
\ref{eq:lambda=00003DDeltaF} differ by up to 25 kJ/mol from the one
computed with the graphical definition. 
\begin{table}
\centering{}{\footnotesize{} }%
\begin{tabular}{|c|c|c|c|c|}
\hline 
{\footnotesize{}$\lambda$} & {\footnotesize{}HNC} & {\footnotesize{}PMV (eq \ref{eq:lambda=00003DDeltaF}) } & {\footnotesize{}PMV graphical} & {\footnotesize{}WDA}\tabularnewline
\hline 
{\footnotesize{}$\text{Cl}$} & {\footnotesize{}297} & {\footnotesize{}332} & {\footnotesize{}287} & {\footnotesize{}295 }\tabularnewline
\hline 
{\footnotesize{}$\text{Cl}^{-}$} & {\footnotesize{}264} & {\footnotesize{}229} & {\footnotesize{}230} & {\footnotesize{}255 }\tabularnewline
\hline 
\hline 
{\footnotesize{}$\text{Cl}$} & {\footnotesize{}218} & {\footnotesize{}220} & {\footnotesize{}220} & {\footnotesize{}209 }\tabularnewline
\hline 
{\footnotesize{}$\text{Cl}^{+}$} & {\footnotesize{}214} & {\footnotesize{}212} & {\footnotesize{}197} & {\footnotesize{}209 }\tabularnewline
\hline 
\end{tabular}\caption{Solvent reorganization free energies (in kJ/mol) calculated with the
HNC functional, with the PMV correction using equation \ref{eq:lambda=00003DDeltaF},
with the PMV correction using the graphical definition and with the
WDA functional.}
\label{tab} 
\end{table}

\begin{figure}[htbp]
\centering{}\includegraphics[width=0.95\columnwidth]{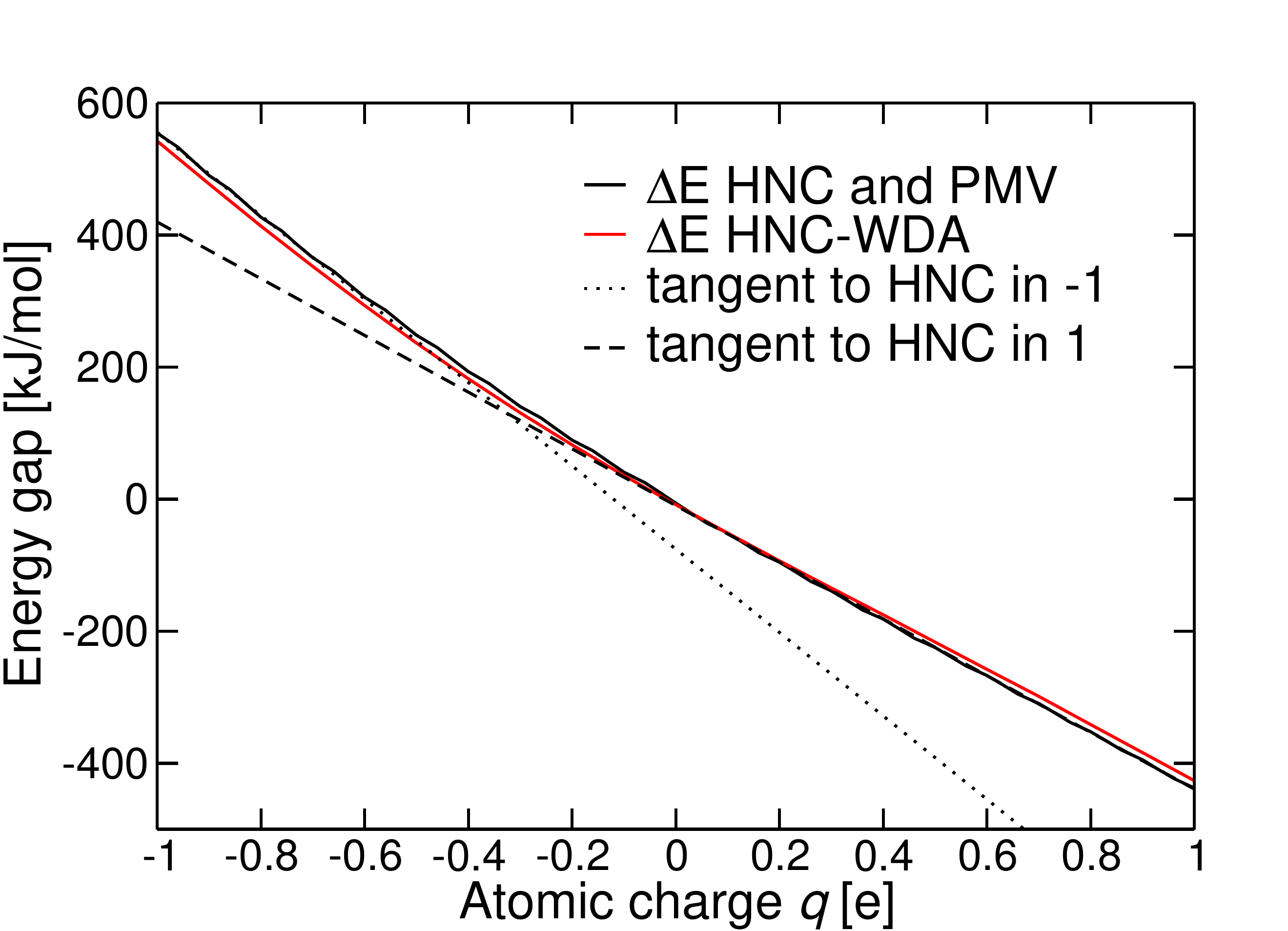}\caption{Vertical energy gap computed using the HNC (solid black) and WDA functional
(solid red) as a function of the atomic charge. The dotted and dashed
lines are the tangent to the HNC curve in $q=-1$ and $q=1$
respectively.}
\label{figdeltaE} 
\end{figure}

The evolution of the energy gap as a function of $q$ is displayed
in figure \ref{figdeltaE}. We recall that values of $q$ between 0 and 1 correspond to the $\text{Cl}\rightarrow\text{Cl}^{+}$
 while the value between 0 and -1 corresponds to $\text{Cl}^{-}\rightarrow\text{Cl}$.
When an \red{electron transfer reaction} is following Marcus theory,
the energy gap should vary linearly. This is the case between $q=0$ and
$q=1$, while the linearity is not respected between $q=0$
and $q=-1$. This is consistent with the conclusions drawn examining
the reorganization free energies computing using the HNC functional
and with the MD results:  the $\text{Cl}\rightarrow\text{Cl}^{+}$
reaction follows Marcus theory while the $\text{Cl}^{-}\rightarrow\text{Cl}$
reaction does not. Since the energy gap is not modified by the PMV correction,
the linear behavior observed for the $\text{Cl}\rightarrow\text{Cl}^{+}$
is in contradiction with the different values of $\lambda$ of table
\ref{tab}. This is another proof of the inappropriateness of the
PMV correction to study \red{electron transfer reactions}. 

\subsection{Weighted density bridge functional}

The same \red{electron transfer reactions} were studied using the WDA functional of equation \ref{eq:FWDA}
and the prediction for the \red{free energy profiles} as a function of the energy gap or
of the solute charge are given in figures \ref{fig1: FEC} and \ref{fig2:FEC_and_PMV_as_q}
respectively. Adding WDA bridge has a limited impact on the \red{free energy profile} as
compared to the HNC predictions. More importantly, the minimum of
the free energy curve of the cation (resp. anion) is correctly reached
for the solute with charge $q=1$ (resp. $q=-1$). This is because
the WDA correction has a functional form and the variational principle
is respected. The reorganization free energies predicted with the
WDA functional are slightly reduced with respect to the one predicted
using HNC but the overall conclusion is recovered: $\text{Cl}\rightarrow\text{Cl}^{+}$
follows Marcus picture while $\text{Cl}\rightarrow\text{Cl}^{-}$
does not. This is also supported by the evolution of the energy gap as
a function of the atomic charge in figure \ref{figdeltaE}.

We now attempt to understand why the reorganization free energy of
the ions are reduced when the WDA bridge is used. We introduce the
free energy difference for a state $\alpha$ as 
\begin{align}
\Delta F_{\alpha}(\eta) & =F_{\alpha}^{\text{WDA}}(\eta)-F_{\alpha}^{\text{HNC}}(\eta)\label{eq:F^func_alpha(eta)}\\%%%
 & =\Delta F_{\alpha}^{\text{ext}}(\eta)+\Delta F_{\alpha}^{\text{int}}(\eta)\label{eq:DF_alpha(eta)}
\end{align}
where $\Delta F_{\alpha}^{\text{ext}}$ is the difference of the external
functionals defined by equation \ref{eq:Fext} and $\Delta F_{\alpha}^{\text{int}}$
is the contribution of the ideal and excess term in equation \ref{F=00003DFid+Fexc+Fext}.
Since the WDA functional modifies the minimization process, the equilibrium solvent densities 
associated to the same $\eta$ differ for both corrections in equation \ref{eq:F^func_alpha(eta)}.

\begin{figure}[htbp]
\centering{}\includegraphics[width=0.95\columnwidth]{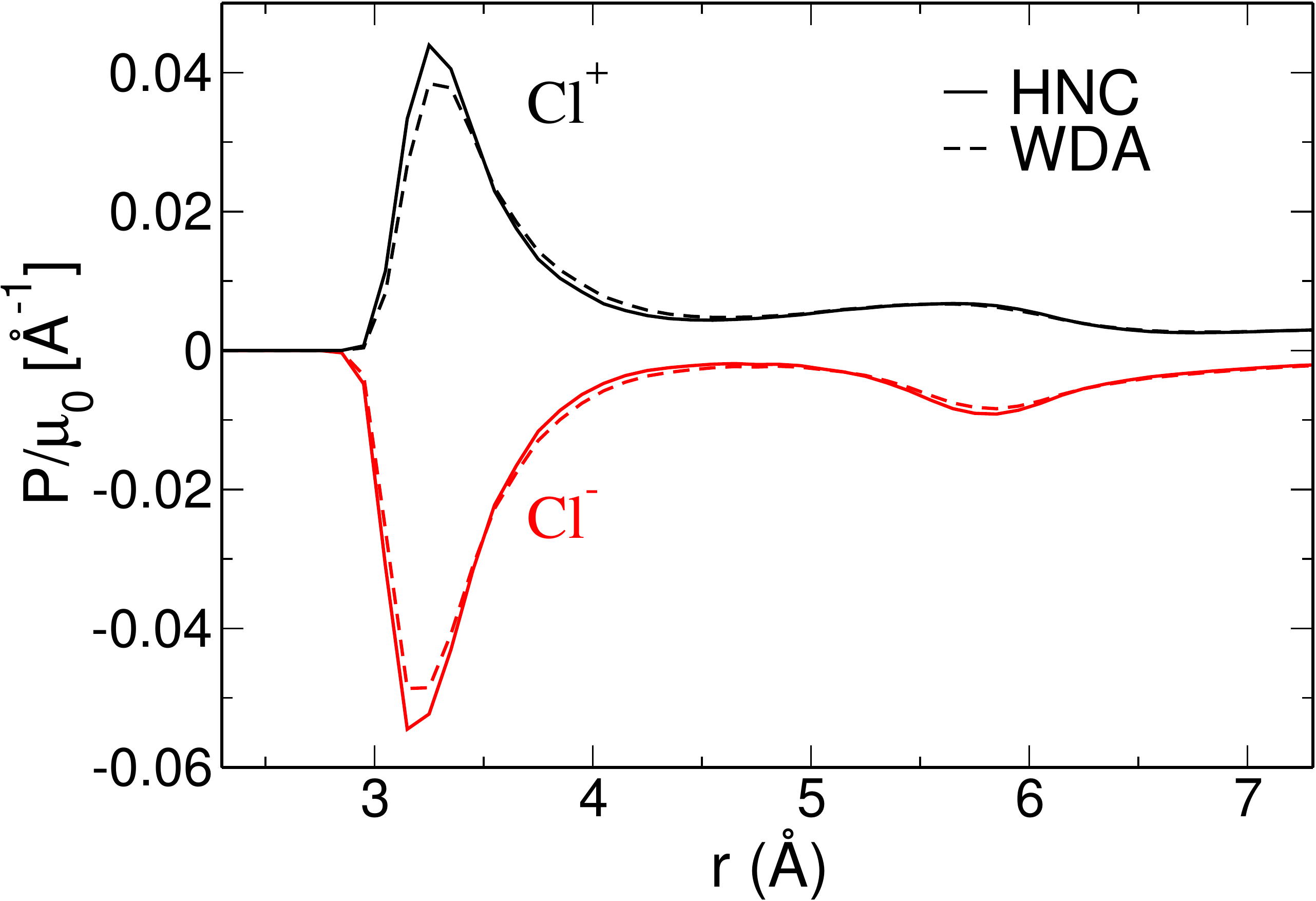}\caption{Equilibrium radial polarization density around $\text{Cl}^{-}$ and
$\text{Cl}^{+}$ computed by HNC and WDA}
\label{figrdf} 
\end{figure}

Using equation
\ref{eq:lambda=00003DDeltaF}, \red{the difference of the reorganization free energies obtained with and without the WDA functional can be expressed as}

\begin{eqnarray}
\Delta\lambda_{\alpha} & = & \red{\lambda_{\alpha}^{\text{WDA}}-  \lambda_{\alpha}^{\text{HNC}}} = \Delta F_{\alpha}(0)-\Delta F_{\alpha}(\alpha)\label{eq:DeltaLambda}\\
 & \approx & \Delta F_{\alpha}^{\text{elec}}(0)-\Delta F_{\alpha}^{\text{elec}}(\alpha)\label{eq:Deltalambdaapprox} %%%
\end{eqnarray}
where $\Delta F_{\alpha}^{\text{elec}}$ is the electrostatic part of external functional difference. %%%
In equation \ref{eq:Deltalambdaapprox}, we assume that since the
considered species are ions, the main contribution is due to the electrostatic,
i.e $\Delta F^{\text{int}}$ and the Lennard-Jones contributions can
be neglected.
As long as we are considering  \red{spherical} ions, the electrostatic potential is spherically symmetric, \red{so is the polarization of the solvent}. %%%
\red{The equilibrium dipolar polarization density of the fluid can be computed from the equilibrium solvent density as 
\begin{equation}
\frac{\bm{P}(\bm{r},\bm{\Omega})}{\mu_0}=\int \rho_{\text{eq}}(\bm{r},\bm{\Omega})\bm{\Omega}d\bm{\Omega}
\end{equation}
where $\mu_0$ is the molecular dipole of the solvent molecule.}
\red{The spherically averaged radial component of the solvent equilibrium polarization
around each ion are displayed in figure \ref{figrdf}. }

The WDA functional reduces the polarization of the solvent in the
vicinity of the charged solutes, as evidenced by the decrease of the
first peaks in figure \ref{figrdf}. Since the polarization is reduced,
the ion is destabilized by the WDA functional when immersed in its
equilibrium density, $\Delta F_{\alpha}^{\text{elec}}(\alpha)>0$.
On the contrary, the equilibrium density of the oppositely charged
ion is less destabilizing $\Delta F_{\alpha}^{\text{elec}}(-\alpha)<0$.
Finally, the WDA correction being angular independent we can expect
the polarization around the neutral solute to be almost unchanged
$\Delta F_{\alpha}^{\text{elec}}(0)\approx0$. With this simple analysis
and using equation \ref{eq:Deltalambdaapprox} we recover the decrease
of the reorganization free energy observed in table \ref{tab}. A
verification of the hypothesis made here and a more detailed examination
of the evolution of the different quantities entering equations \ref{eq:DF_alpha(eta)}-\ref{eq:DeltaLambda}
is available in SI.

\section{Conclusions}

In this paper we studied the appropriateness of two corrections to
the HNC functional to study aqueous electron transfer reactions with
molecular density functional theory. First, we examine a simple a
posteriori correction of the pressure which has been widely used with
different expression in mDFT and 3D-RISM\cite{sergiievskyi_fast_2014,gillespie_restoring_2014,sergiievskyi_solvation_2015,misin_communication_2015,misin_salting-out_2016,robert_pressure_2020,luukkonen_hydration_2020}.
Despite its success to predict solvation free energy in good agreement
with reference simulations and experiments, this correction should
not be used to study \red{electron transfer reactions}. The minima of the \red{free energy profile} does not correspond
to the equilibrium solvent configuration when this correction is used
because it modifies the free energy without affecting the functional
optimization. The reorganization free energies becomes ill-defined:
the graphical definition do not coincide with the mathematical expression.
Moreover, we have shown that using any of the two definitions of the
reorganization free energy lead to a behavior deviating from the Marcus
picture for $\text{Cl}\rightarrow\text{Cl}^{+}$ reaction in disagreement
with MD simulations and HNC mDFT calculations. This deviation form
Marcus theory is not recovered when examining the evolution of the
energy gap, which is a final evidence for the inconsistency of the
pressure correction.

We then turned to another type of correction, a so-called bridge functional
which is trying to recover some of the contribution due to the terms
beyond the quadratic approximation of the HNC functional. We have
chosen to use the recent and simple angular independent weighted density
functional \cite{borgis_simple_2020} that was shown to properly reproduce
the solvation free energies of hydrophobic solutes. Since this bridge
has a functional form, the equilibrium density is modified and this
approach does not suffer from the flaws of the PMV correction. There
are no ambiguities in the definitions of the reorganization free energies,
and we recover results consistent with MD simulations and HNC mDFT
calculations: $\text{Cl}\rightarrow\text{Cl}^{+}$ follows Marcus
picture while $\text{Cl}^{-}\rightarrow\text{Cl}$ deviates from it.
Overall, the WDA functional does not modify significantly the results
obtained without correction \cite{jeanmairet_molecular_2019}. This
might be because it is an angular independent correction, having a
low impact on the polarization of the solvent which is the dominant
effect for ions. 

\section{Acknowledgement}
The authors acknowledge Mathieu Salanne for his constructive comments about the manuscript.
This work has been supported by the Agence Nationale de la Recherche, projet ANR BRIDGE AAP CE29.

\section{Data Availability Statement}
Data available on request from the authors

\bibliographystyle{plain}

 \bibliographystyle{unsrt} 
\end{document}